\documentclass[conference]{IEEEtran}
\IEEEoverridecommandlockouts
\usepackage{amsmath,amssymb,amsfonts}
\usepackage{algorithmic}
\usepackage{graphicx}
\usepackage{textcomp}
\usepackage{xcolor}
\usepackage{hyperref}
\usepackage[sorting=none]{biblatex}
\bibliography{bibs}
\begin{document}

\title{Evaluating the Effect of Compression on Video Temporal Consistency Using Objective Quality Metrics\\

\thanks{The author thanks the EMJM Imaging program for their valuable support.}
}

\author{\IEEEauthorblockN{Peter Zsoldos}
\IEEEauthorblockA{\textit{Sensible Things that Communicate} \\
\textit{Mid Sweden University}\\
Sundsvall, Sweden \\
zsoldosp44@gmail.com}

}
\maketitle
\begin{abstract}
While video compression algorithms effectively reduce bitrate, aggressive quantization often compromises temporal coherence, introducing artifacts such as flicker, motion inconsistency, and unstable textures. Although spatial quality degradation is well-documented, the relationship between compression intensity and temporal stability remains insufficiently characterized. This paper systematically examines the progression of frame-to-frame coherence errors across different bitrate regimes, utilizing multiple codecs (AV1, HEVC, VP9, H.264) and content types. Our findings reveal that temporal consistency degrades non-linearly with increasing compression. Most critically, we identify a ``Predictability anomaly'' where sequences with unpredictable or irregular dynamics experience disproportionately higher instability than sequences with higher, but more predictable, motion magnitude. This challenges the conventional assumption that motion volume alone dictates encoding difficulty and highlights the necessity of temporal-aware metrics in compression pipelines.
\end{abstract}

\section{Introduction}

Video compression is fundamental to modern streaming and real-time communication, enabling efficient transmission by removing spatial and temporal redundancies. While modern codecs achieve impressive bitrate reductions, they inevitably introduce distortions that manifest over time. Among these, temporal inconsistencies—such as frame-to-frame flicker, ``floating'' textures, and motion jitter—are critically important yet often overlooked by traditional spatial quality assessments.

Existing research has extensively characterized spatial artifacts like blocking and blurring. However, most current assessments rely on frame-by-frame metrics like PSNR, which fail to capture the dynamic lack of coherence that severely impacts human perceptual quality in motion-heavy content. To bridge this gap, we distinguish "temporal consistency"—the preservation of visual features like texture and brightness over time—from "temporal stability," which we define as the encoder's ability to maintain a reliable inter-frame prediction loop without reverting to frequent I-frame "resets" that cause flickering or "breathing" artifacts . This work addresses the resulting research gap by systematically analyzing how varying compression levels influence these temporal properties across four codec generations: H.264, HEVC, VP9, and AV1. 

By utilizing a suite of objective temporal metrics, we quantify the onset of artifacts and identify a critical non-linear relationship between content complexity and stability. Uniquely, our findings challenge the assumption that motion volume is the primary driver of compression difficulty; we reveal that \textit{unpredictable} dynamics cause significantly greater instability than high-magnitude, structured motion.

\section{Related Work}

The impact of video compression on temporal coherence has been studied through artifact detection, quality assessment, and post-processing~\cite{jimenez2014standard}. Lossy compression introduces distinct temporal artifacts, such as flickering and ``floating'' textures, which traditional frame-by-frame metrics (e.g., PSNR, SSIM) fail to capture~\cite{lin2024toward}. To address this, Lin \textit{et al.} developed the PEA265 database and SSTAM metric, demonstrating the perceptual significance of these spatio-temporal distortions~\cite{Lin2019PEA265, Lin2023SSTAM}. 

Recognizing these limitations, researchers have proposed metrics that integrate temporal information, ranging from the Spatio-Temporal SSIM (ST-SSIM)~\cite{Wang2025STSSIM} and motion-compensated filtering~\cite{Song2017MCTF} to recent two-stream deep learning networks that explicitly model temporal degradation~\cite{He2024TSCNN}. While frameworks like Q-STAR have modeled the trade-offs between resolution and quantization~\cite{Ou2012QSTAR}, they often overlook frame-to-frame coherence under bitrate-driven constraints. This leaves a critical gap in understanding how compression intensity specifically influences temporal stability, motivating the systematic investigation presented in this study.

This study aims to answer the following research question: How do successive generations of video codecs (H.264 through AV1) balance bitrate efficiency with temporal stability, and to what extent does motion predictability, rather than simple motion volume—dictate the onset of temporal artifacts and the reliability of objective quality metrics?

\section{Methodology}
The code for this paper can be reached at: https://github.com/Hannibal0319/video\_compression

\subsection{Datasets and Complexity Grouping}
We utilized two benchmark datasets: the Ultra Video Group (UVG) dataset (1080p, 50fps)~\cite{UVG2018}, the HEVC Class B dataset (1080p, 24--60fps)~\cite{Sullivan2012HEVC} and the BVI-HD dataset (1080p, 50–120fps)~\cite{BVI}, providing a wider variety of textures and motion patterns. From these sources, a total of 44 aggregated sequences were selected for evaluation. To systematically evaluate motion dynamics, we categorize the sequences using \textbf{Temporal Information (TI)}. TI is defined as the maximum standard deviation over time of the space-time difference between consecutive frames, representing the "dynamic energy" or motion volume within a sequence. 

The 44 aggregated sequences were sorted by their TI values and divided into four quartiles to analyze the impact of different motion types:
\begin{itemize}
    \item \textbf{TI Group 1 (Static):} 0--25\textsuperscript{th} percentile; contains content with stationary backgrounds and minimal movement.
    \item \textbf{TI Group 2 (Predictable Motion):} 25--50\textsuperscript{th} percentile; low-to-medium motion with high temporal structure.
    \item \textbf{TI Group 3 (Unpredictable Dynamics):} 50--75\textsuperscript{th} percentile; medium-to-high irregular motion (e.g., water, crowds) where motion vectors lack clear structure.
    \item \textbf{TI Group 4 (Global Motion):} 75--100\textsuperscript{th} percentile; high-magnitude but structured motion, such as consistent camera pans.
\end{itemize} This stratification ensures a balanced evaluation across diverse motion characteristics.

\subsection{Codecs and Compression Settings}
We evaluated four codecs representing different generations: H.264, HEVC, VP9, and AV1. VVC was left out, because restricting its bitrate, which is crucial for fair comparison, prevents it from using its strongest RDO and partitioning behavior. This is not its intended use. 
All encodings were performed using FFmpeg libraries with parameters optimized for perceptual quality while constraining rate control. A summary of the encoder configurations is provided in Table~\ref{tab:codecs}.
\begin{table}[htbp]
\centering
\caption{Codec Configurations and Encoder Settings}
\label{tab:codecs}
\begin{tabular}{l l l}
\hline
\textbf{Codec} & \textbf{Encoder} & \textbf{Key Parameters} \\
\hline
H.264 & \texttt{libx264} & \texttt{slow}, 8-bit, \texttt{faststart}\\
HEVC & \texttt{libx265} & \texttt{slow}, 10-bit, \texttt{aq-mode=3} \\
VP9 & \texttt{libvpx-vp9} & \texttt{row-mt 1}\\
AV1 & \texttt{libsvtav1} & \texttt{preset 4}, \texttt{rc=1} \\
\hline
\end{tabular}
\end{table}

Videos were encoded at 7 compression levels and the target bitrate was defined as $\text{Bitrate} = L \times 1000~\text{kbps}$, where $L$ represents the scaling factor i.e. the compression level. The levels computed were: 1, 1.5, 2, 2.5, 3, 4 and 8.

\subsection{Evaluation Metrics}
To quantify the impact of compression, we employed a comprehensive set of metrics:
\begin{itemize}
    \item \textbf{Spatial Quality:} PSNR, SSIM, and VMAF~\cite{VMAF2016}.
    \item \textbf{Temporal Consistency:} This study utilizes metrics that explicitly model inter-frame dynamics. ST-RRED~\cite{ST_RRED2011} calculates the fluctuation of information content across space and time to detect instabilities missed by spatial-only assessments. The MOVIE Index evaluates quality along motion trajectories, identifying distortions like "floating" textures by tracking how errors manifest along the path of moving objects. Additionally, T-SSIM and T-PSNR are used to measure structural and signal-level coherence between consecutive frames.
    \item \textbf{Generative Quality:} Fréchet Video Distance (FVD)~\cite{FVD2018} was used to measure feature distribution discrepancies.
    \item \textbf{Rate-Distortion Efficiency:} Bjøntegaard Delta (BD) Rate curves were computed to quantify the average percentage difference in bitrate required to achieve the same objective quality between the tested codecs and the anchor.
\end{itemize}

\section{Results}
\subsection{Objective Performance Hierarchy}
\begin{figure*}[htbp]
    \centering
    \includegraphics[width=0.9\textwidth]{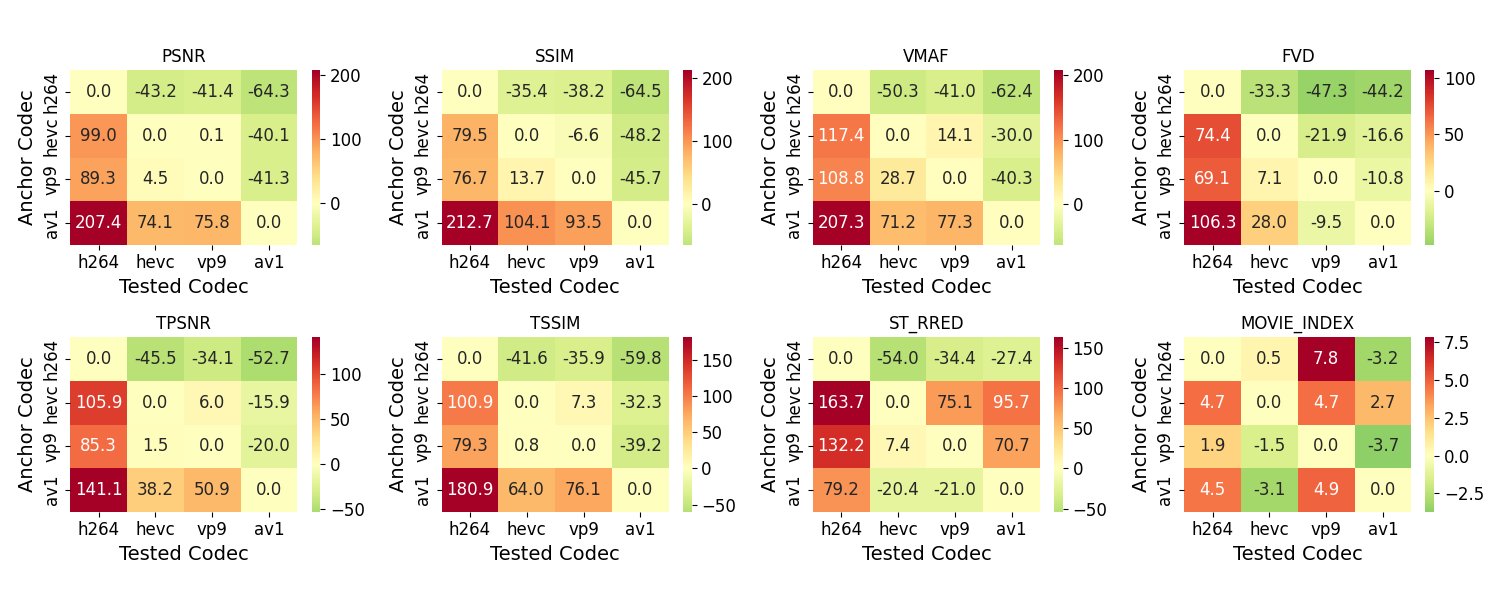}
    \caption{Average BD-Rate Scores - BVI-HD Dataset. The matrices compare the Bjøntegaard Delta Rate (BD-Rate) of tested codecs against anchor codecs across spatial (PSNR, SSIM, VMAF) and temporal (FVD, TSSIM, MOVIE\_INDEX) metrics. Negative values (green) indicate bitrate savings (better performance), while positive values (red) indicate bitrate increases (worse performance) to achieve the same quality. MOVIE\_INDEX is shown with a logarithmic scale for better visualization.}
    \label{fig:bd_rate_scores}
\end{figure*}

The objective analysis confirms a clear generational performance hierarchy across all tested codecs. As illustrated in the BD-Rate comparisons (Fig.~\ref{fig:bd_rate_scores}), \textbf{AV1} consistently yields the highest Rate-Distortion (R-D) performance, offering the best quality for a given bitrate. \textbf{HEVC} follows closely, tracking AV1's performance and significantly outperforming legacy standards. While \textbf{VP9} generally provides superior efficiency compared to \textbf{H.264}, the performance gap between them narrows at higher bitrates.

\subsection{Content-Based Analysis and the TI Anomaly}
Typically, objective quality is expected to decrease as TI increases. However, our analysis reveals a significant "Predictability Anomaly" between the highest complexity groups.

\paragraph{The Fallback to Intra-Coding}As shown in Table~\ref{tab:iframes}, \textbf{TI Group 2} and \textbf{TI Group 3} exhibit the highest average I-frame frequencies (0.61\% and 0.57\% respectively). This indicates that the codecs struggle to establish reliable motion vectors for these sequences, forcing the encoder to "reset" the prediction loop more frequently. While Group 4 has higher raw motion magnitude, its structured nature (e.g., global pans) allows for better inter-frame prediction, resulting in the lowest I-frame frequency (0.50\%).

\paragraph{The VMAF-Temporal Paradox}Interestingly, this inability to utilize temporal redundancy in Group 3 results in a unique behavior at higher bitrates ($\geq$ 4000 kbps). In this regime, \textbf{VMAF scores for TI Group 3 surpass all other groups.} Because the encoder is forced into a heavy I-frame strategy, it essentially produces a sequence of high-fidelity spatial snapshots. Since VMAF is heavily weighted toward spatial features like Visual Information Fidelity (VIF) and Detail Loss Measure (DLM), it rewards this spatial sharpness, even though the coding is temporally inefficient.

\begin{table}[htbp]
\centering
\caption{Average Percentage of I-Frames Produced by Videos from Different TI Groups}
\label{tab:iframes}
\begin{tabular}{cc}
\hline
\textbf{TI Group} & \textbf{Average I-Frame \%} \\
\hline
2 & 0.6146\% \\
3 & 0.5694\% \\
1 & 0.5417\% \\
4 & 0.5000\% \\
\hline
\end{tabular}
\end{table}
\begin{figure}
    \centering
    \includegraphics[width=1\linewidth]{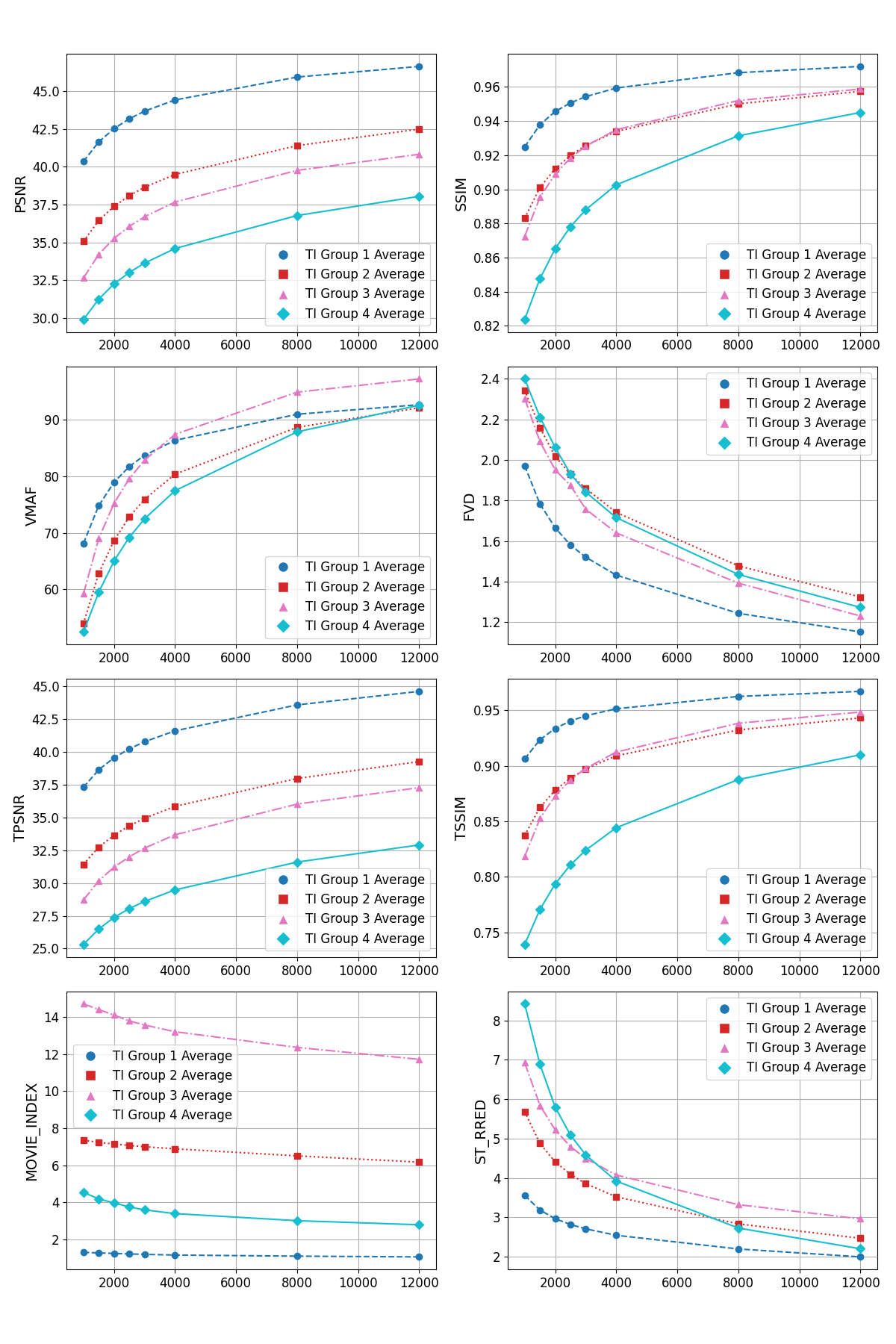}
    \caption{Rate-Distortion (R-D) curves comparing objective spatial and temporal quality metrics across four content complexity groups (TI Groups 1--4) averaged across videos per group and codecs. The plots display quality scores ($y$-axis) as a function of bitrate ($x$-axis) for various metrics, including VMAF, PSNR, SSIM, and FVD.  The results highlight the ``TI Group Anomaly,'' where the unpredictable dynamics of \textbf{TI Group 3} consistently result in lower quality performance than the higher-magnitude but more regular motion of \textbf{TI Group 4}. FVD score is shown with a logarithmic scale for better visualization.}
    \label{fig:TI_groups}
\end{figure}

\begin{figure}[htbp]
    \centering
    \includegraphics[width=\linewidth]{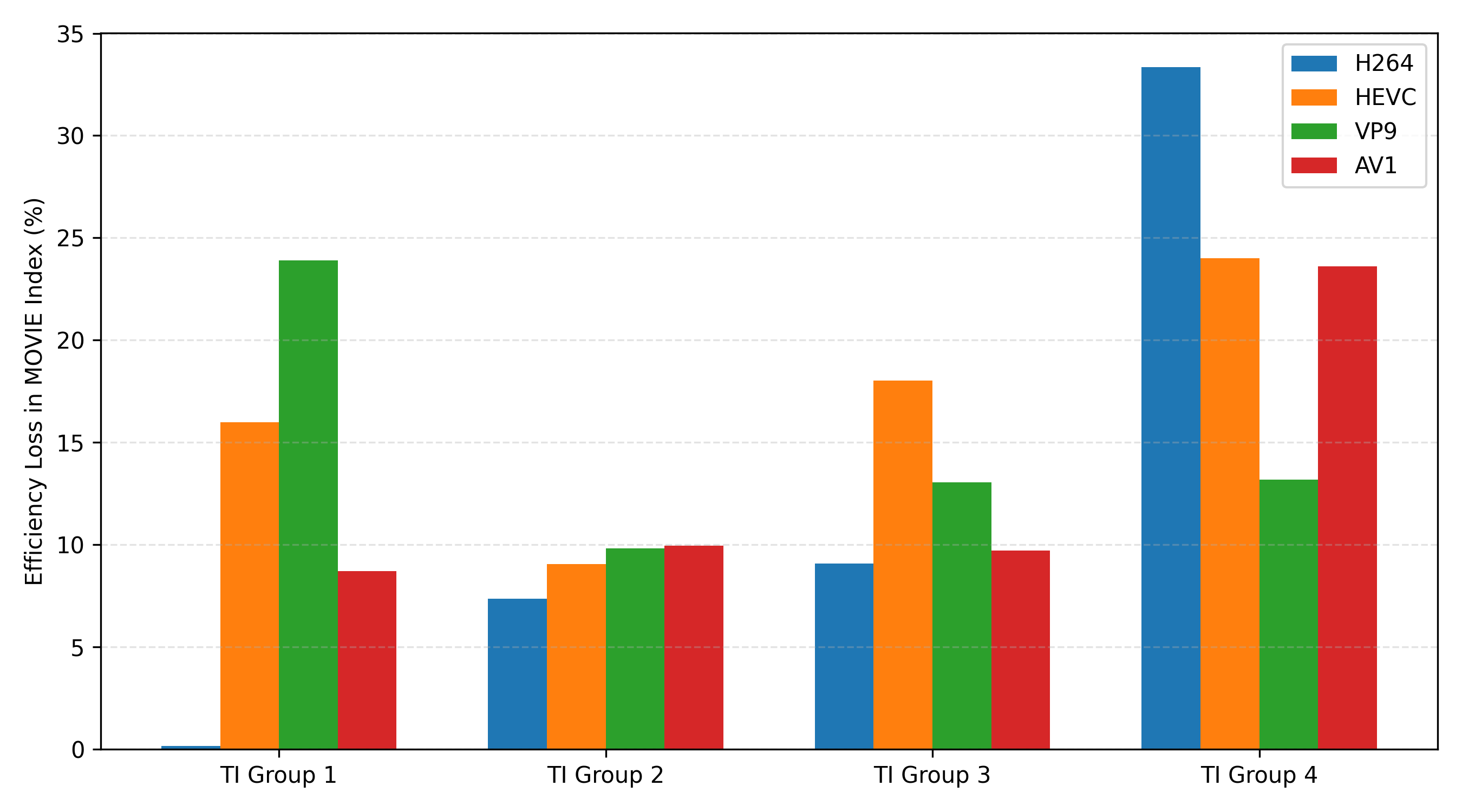}
    \caption{Relative Temporal Stability Recovery ($2000 \rightarrow 8000$ kbps). This plot illustrates the percentage of total temporal distortion (MOVIE Index) removed when quadrupling the bitrate. While all groups show absolute improvement, the relative recovery for TI Group 4 (Global Motion) is significantly higher than that of TI Group 3 (Unpredictable Dynamics) across all codecs, highlighting the "stubborn" nature of irregular motion.}
    \label{fig:movie_reduction}
\end{figure}

\begin{figure}[htbp]
    \centering
    \includegraphics[width=\linewidth]{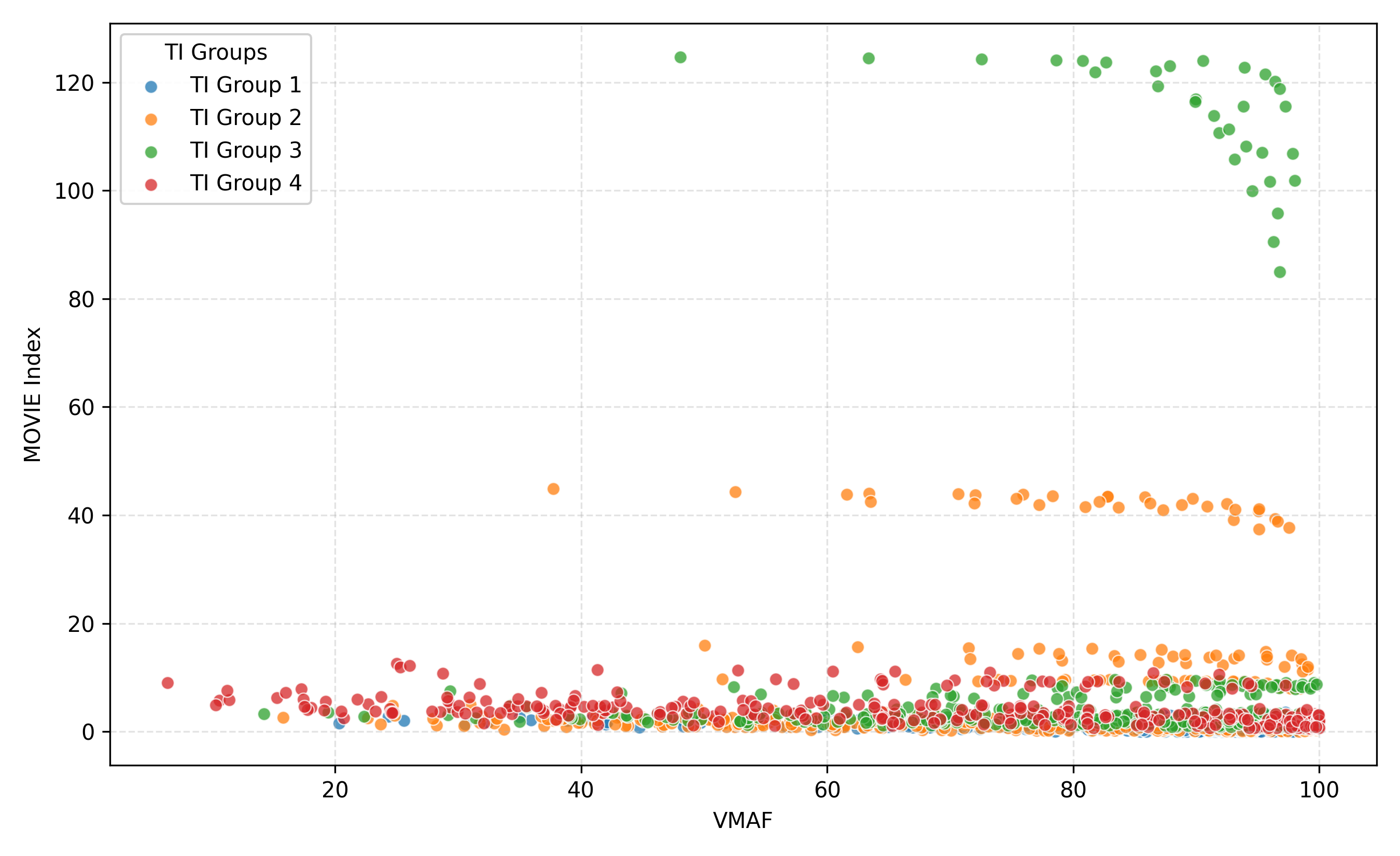}
    \caption{Correlation between Spatial Quality (VMAF) and Temporal Distortion (MOVIE Index). TI Group 3 (green) forms a distinct "floating" cluster, showing high temporal distortion even at near-perfect VMAF scores.}
    \label{fig:vmaf_movie}
\end{figure}

\begin{figure}[htbp]
    \centering
    \includegraphics[width=\linewidth]{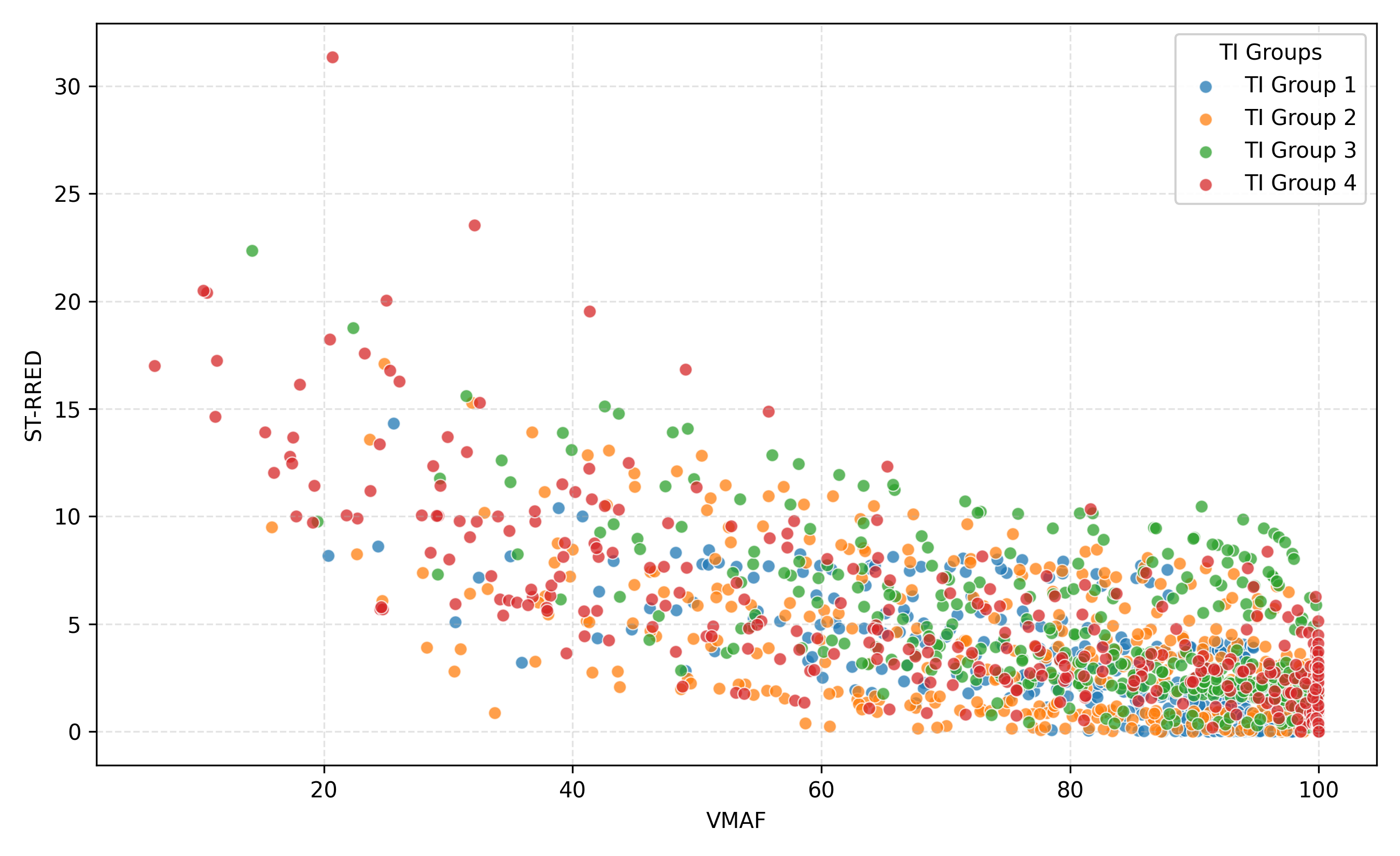}
    \caption{Correlation between VMAF and Information Fluctuation (ST-RRED). Unlike Groups 1, 2, and 4, which converge toward stability at high bitrates, TI Group 3 maintains high error variance regardless of spatial quality.}
    \label{fig:vmaf_ST-RRED}
\end{figure}

\subsection{Metric Sensitivity}
\begin{itemize} \item \textbf{VMAF Paradox:} At higher bitrates, VMAF scores for TI Group 3 surpass all others. This is likely a byproduct of the high I-frame frequency; as bitrates increase, these frequent I-frames become nearly lossless spatial anchors. Since VMAF is heavily weighted toward spatial features (DLM and VIF), it rewards this spatial clarity. \item \textbf{Temporal Distortion (ST-RRED \& MOVIE):} In contrast, \textbf{ST-RRED} and \textbf{MOVIE Index} show that TI Group 3 remains highly distinct and problematic. \textbf{ST-RRED} measures the fluctuation of information content across space and time, while the \textbf{MOVIE Index} evaluates distortions along motion trajectories. These metrics reveal that while the "I-frame heavy" strategy fixes spatial quality, it fails to solve—and may even exacerbate—temporal jitter and "breathing" artifacts that occur when the prediction loop is frequently reset. \end{itemize}

The visual evidence for the TI Group 3 anomaly is most apparent when examining the relative efficiency of quality recovery, as shown in Fig. \ref{fig:movie_reduction}. While Group 3 exhibits large absolute reductions in distortion due to its high initial error state, its percentage of distortion removed is consistently lower than that of Group 4. For example, in H.264, Group 4 sees a stability improvement of over 30\%, while Group 3 remains below 10\%. This disparity confirms that encoders are less efficient at stabilizing unpredictable motion paths, even when provided with a 400\% increase in bitrate. This lack of relative recovery is the 'smoking gun' for the VMAF Paradox: the encoder uses the extra bits to sharpen individual frames (improving VMAF), but cannot successfully resolve the underlying motion-path inconsistencies (leaving the MOVIE Index relatively high).

The profound decoupling between spatial and temporal metrics is visually isolated in Fig. \ref{fig:vmaf_movie} and Fig. \ref{fig:vmaf_ST-RRED}. While TI Groups 1, 2, and 4 exhibit a clustering trend where high spatial fidelity correlates with lower temporal error, TI Group 3 (Unpredictable Dynamics) acts as a persistent outlier.

In the VMAF vs. MOVIE Index plot (Fig. \ref{fig:vmaf_movie}), Group 3 sequences reach VMAF scores exceeding 90 while simultaneously exhibiting the highest distortion values in the entire dataset. Although one might note that all groups exhibit scatter, the nature of this scatter is distinct: Groups 1, 2, and 4 'funnel' toward stability at high bitrates, whereas Group 3 maintains a high error variance that is absent in structured motion groups like TI Group 4. This confirms that encoders effectively 'cheat' spatial metrics by prioritizing frame-by-frame sharpness through I-frame 'resets' (Table II) at the expense of motion-trajectory integrity.
\section{Discussion}

The Rate-Distortion (R-D) analysis across the four codecs (AV1, HEVC, VP9, H.264) and four content complexity groups (TI Groups 1-4) confirms expected generational performance gains while simultaneously highlighting the profound and non-linear impact of content characteristics on compression efficiency.

\subsection{Generational Performance and Efficiency}

As expected, the modern codecs, particularly \textbf{AV1}, consistently demonstrate superior efficiency, yielding higher objective quality (VMAF, PSNR) across the entire bitrate range compared to the legacy codecs H.264 and VP9. This performance differential underscores the advancements in motion compensation, spatial prediction, and transform coding implemented in the newer standards. The close tracking between AV1 and HEVC suggests that while AV1 achieves marginal gains, HEVC remains a highly competitive standard, particularly in constrained environments.

\subsection{The Non-Linear Impact of Content Complexity}

The most significant finding is the non-linear relationship between the quantitative measure of complexity (TI) and encoding performance.

\paragraph{TI Group 1: The Ideal Case}
Content with the lowest complexity (TI Group 1) predictably sets the performance ceiling, affirming that static or low-motion content remains the easiest to compress.

\paragraph{The TI Group Anomaly}
The correlation between high I-frame counts and high high-bitrate VMAF scores suggests a "fallback strategy" in modern codecs. When faced with the unpredictable dynamics of TI Group 3, the motion estimation engine effectively "gives up," resulting in higher I-frame insertion. While this is inefficient for BD-rate, it results in a sequence of high-quality still images. This explains why temporal metrics (ST-RRED, MOVIE) continue to flag these sequences as unstable even when spatial metrics (VMAF) suggest near-perfect quality.
As visualized in Fig. \ref{fig:movie_reduction}, there is a clear efficiency ceiling for TI Group 3. The unpredictable nature of sequences like crowds or moving water lacks the temporal structure necessary for inter-frame prediction to thrive, regardless of bitrate. In contrast, TI Group 4—despite having higher raw motion magnitude—shows the highest percentage of quality recovery, proving that motion regularity is a more significant determinant of encoder success than motion volume.

\subsection{The Role of Quality Metrics}

The analysis using multiple quality metrics provides crucial context:
\begin{itemize}
    \item \textbf{VMAF} consistently shows a greater ability to differentiate between the TI groups compared to traditional PSNR, supporting its relevance as a perceptually aligned metric for contemporary video evaluation.
    \item The temporal metrics, \textbf{FVD} and \textbf{ST-RRED}, confirm that temporal quality degrades most severely for the highest complexity groups at low bitrates, but also show that modern codecs (AV1, VP9) are more effective at mitigating this temporal degradation at higher bitrates.
\end{itemize}

\section{Conclusion}
This study systematically evaluated the progression of temporal consistency across four codec generations, identifying a critical "complexity anomaly" that challenges traditional encoding assumptions. Our findings reveal that motion regularity is a more significant determinant of temporal stability than raw motion magnitude.

\begin{itemize}
    \item \textbf{The Predictability Paradox:} While high-magnitude but structured motion (Global Motion) is handled efficiently through inter-frame prediction, unpredictable dynamics and complex textures force encoders into a high-frequency I-frame fallback strategy.
    \item \textbf{Metric Limitations:} This strategy creates a "VMAF Paradox" at high bitrates, where spatial-heavy metrics like VMAF yield artificially high scores by rewarding a sequence of sharp, nearly lossless spatial snapshots.
    \item \textbf{Temporal Instability:} In contrast, temporal-aware metrics such as ST-RRED and MOVIE Index expose a persistent lack of coherence and motion-path integrity in these sequences, even when spatial scores suggest near-perfect quality.
\end{itemize}

Ultimately, this research highlights a significant gap in current evaluation pipelines: encoders can effectively "cheat" spatial metrics by sacrificing temporal efficiency for frame-by-frame sharpness. Future compression standards must prioritize motion-trajectory consistency in rate-distortion optimization to better manage unpredictable content types

\section{Future Work}
To further investigate the root causes of the TI Group 3 anomaly, future research will focus on:
\begin{itemize}
    \item \textbf{Content Characterization:} Analyzing specific spatial features (e.g., texture density, noise levels) to pinpoint the exact characteristics that disrupt motion estimation.
    \item \textbf{Encoder Tool Analysis:} Examining internal codec statistics, such as motion vector distribution and mode decision frequencies, to understand which specific prediction tools fail when encountering unpredictable dynamics.
    \item \textbf{Subjective Validation:} Conducting a Mean Opinion Score study to measure how the anomaly affects the Quality of Experience.
\end{itemize}
\newpage
\printbibliography

\end{document}